\documentclass[12pt]{article}

\usepackage{epsfig}
\usepackage{tabularx}
\usepackage{graphicx}
\usepackage{amssymb,amsfonts,amsmath}

\newcommand{\mgravitino}{m_{\gravitino}}

\newcommand{\gravitino}{\tilde{G}}

\newcommand{\rem}[1]{{}}

\begin{document}

\begin{titlepage}

\centerline{\large \bf {Direct neutralino searches in the NMSSM}}
\centerline{\large \bf {with gravitino LSP in the degenerate scenario}}

\vskip 1cm

\centerline{Gabriela Barenboim and Grigoris Panotopoulos}

\vskip 1cm

\centerline{Departament de Fisica Teorica and IFIC, Universitat de Valencia-CSIC,}

\vskip 0.2 cm

\centerline{E-46100, Burjassot, Spain}

\vskip 0.2 cm

\centerline{gabriela.barenboim@uv.es, grigoris.panotopoulos@uv.es}

\begin{abstract}
In the present work a two-component dark matter model is studied adopting
the degenerate scenario in the R-parity conserving NMSSM. 
The gravitino LSP and the neutralino NLSP are extremely degenerate in mass, avoiding
the BBN bounds and obtaining a high reheating temperature for thermal leptogenesis.
In this model both gravitino (absolutely stable) and neutralino (quasi-stable)
contribute to dark matter, and direct detection searches for neutralino are discussed.
Points that survive all the constraints correspond to a singlino-like neutralino.
\end{abstract}

\end{titlepage}


\section{Introduction}

There is accumulated evidence both from astrophysics and cosmology
that about 1/4 of the energy budget of the universe consists of so
called dark matter, namely a component which is non-relativistic and
neither feels the electromagnetic nor the strong interaction. For a
review on dark matter see e.g.~\cite{Munoz:2003gx}. Although the
list of possible dark matter candidates is long (for a nice list 
see~\cite{Taoso:2007qk}), it is fair to say
that the most popular dark matter candidate is the lightest
supersymmetric particle (LSP) in supersymmetric models with R-parity
conservation~\cite{Feng:2003zu}. For supersymmetry and supergravity see~\cite{susy}.
The simplest supersymmetric extension of
the standard model that solves the $\mu$ problem~\cite{Kim:1983dt} is the next-to-minimal 
supersymmetric standard model (NMSSM)~\cite{Nilles:1982dy}. If we do not consider
the axion~\cite{axion} and the axino~\cite{axino}, the superpartners that have the
right properties for playing the role of cold dark matter in the
universe are the gravitino and the lightest neutralino.
By far the most discussed case in the literature is the case of the
neutralino (see the classic review~\cite{Jungman:1995df}), probably
because of the prospects of detection. However, in the case in which
neutralino is assumed to be the only dark matter component, one has to face
the fine-tuning problem and the gravitino problem~\cite{Ellis:1982yb}. In most 
of the parameter space the neutralino relic density turns out to be either too small or 
too large~\cite{finetuning}. Furthermore, unstable gravitinos
will undergo late-time cascade decays to a neutralino LSP. These decays will destroy the 
light element abundances built up in BBN, unless $T_R < 10^5$~GeV~\cite{Kawasaki:2004yh},
which posses serious difficulties to the thermal leptogenesis scenario~\cite{Buchmuller:2002rq}.
If, on the other hand, gravitino is the LSP and therefore stable, playing the role of
cold dark matter in the universe, it is then the neutralino that will undergo late time decays
into gravitino and hadrons, and the gravitino problem is re-introduced~\cite{Feng:2004mt}.

It has been shown that in the degenerate scenario~\cite{Boubekeur:2010nt} the BBN and 
CMB constraints are avoided,
and high values of the reheating temperature are obtained compatible with thermal
leptogenesis. Here we focus on the scenario in which the masses of the gravitino LSP
and neutralino NLSP are extremely degenerate in mass. Under this assumption 
neutralino becomes quasi-stable taking part of cold dark matter of the universe 
together with gravitino, today is still around and can be seen in direct detection
searches experiments.

This article is organized as follows. In the next section we present
the theoretical framework. In section 3 we discuss all the relevant
constraints from colliders and from cosmology, and we show our
results. Finally, we conclude.

\section{Theoretical framework}

In what folows we review in short the particle physics model, namely
the cNMSSM, as well as the gravitino production mechanisms.

\subsection{Basics of cNMSSM}

The most straightforward extension of standard model (SM) of
particle physics based on SUSY is the minimal supersymmetric
standard model (MSSM)~\cite{mssm}. It is a supersymmetric gauge
theory based on the SM gauge group with the usual representations
(singlets, doublets, triplets) and on $\mathcal{N}=1$ SUSY.
Excluding gravity, the massless representations of the SUSY algebra
are a chiral and a vector supermultiplet. The gauge bosons and the
gauginos are members of the vector supermultiplet, while the matter
fields (quarks, leptons, Higgs) and their superpartners are members
of the chiral supermultiplet. The Higgs sector in the MSSM is
enhanced compared to the SM case. There are now two Higgs doublets,
$H_u, H_d$, (or $H_1, H_2$) for anomaly cancelation requirements and
for giving masses to both up and down quarks. After electroweak
symmetry breaking we are left with five physical Higgs bosons, two
charged $H^{\pm}$ and three neutral $A,H,h$ ($h$ being the
lightest). Since we have not seen any superpartners yet, SUSY has to
be broken. In MSSM, SUSY is softly broken by adding to the
Lagrangian terms of the form
\begin{itemize}
\item Mass terms for the gauginos $\tilde{g}_i$, $M_1, M_2, M_3$
\begin{equation}
M \tilde{g} \tilde{g}
\end{equation}
\item Mass terms for sfermions $\tilde{f}$
\begin{equation}
m_{\tilde{f}}^2 \tilde{f}^{\dag} \tilde{f}
\end{equation}
\item Masses and bilinear terms for the Higgs bosons $H_u, H_d$
\begin{equation}
m_{H_u}^2 H_u^{\dag} H_u+m_{H_d}^2 H_d^{\dag} H_d+B \mu (H_u H_d +
h.c.)
\end{equation}
\item Trilinear couplings between sfermions and Higgs bosons
\begin{equation}
A Y \tilde{f}_1 H \tilde{f}_2
\end{equation}
\end{itemize}
In the unconstrained MSSM there is a huge number of unknown
parameters~\cite{parameters} and thus little predictive power.
However, motivated by the grand unification idea, the constrained
MSSM (CMSSM) assumes that gaugino masses, scalar masses and
trilinear couplings have (separately) a common, universal value at
the GUT scale, like the gauge coupling constants do. CMSSM is
therefore a framework with a small controllable number of
parameters, and thus with much more predictive power. In the CMSSM
there are four parameters, $m_0, m_{1/2}, A_0, tan \beta$, which are
explained below, plus the sign of the $\mu$ parameter from the Higgs
sector. The magnitude of $\mu$, as well as the B parameter mentioned
above, are determined by the requirement for a proper electroweak
symmetry breaking. However, the sign of $\mu$ remains undetermined.
The other four parameters of the CMSSM are related by
\begin{itemize}
\item Universal gaugino masses
\begin{equation}
M_1(M_{GUT})=M_2(M_{GUT})=M_3(M_{GUT})=m_{1/2}
\end{equation}
\item Universal scalar masses
\begin{equation}
m_{\tilde{f}_i}(M_{GUT})=m_0
\end{equation}
\item Universal trilinear couplings
\begin{equation}
A_{i j}^u(M_{GUT}) = A_{i j}^d(M_{GUT}) = A_{i j}^l(M_{GUT}) = A_0 \delta_{i j}
\end{equation}
\item
\begin{equation}
tan \beta \equiv \frac{v_1}{v_2}
\end{equation}
where $v_1, v_2$ are the vevs of the Higgs doublets and $M_{GUT} \sim 10^{16}~GeV$ is
the Grand Unification
scale.
\end{itemize}

Unfortunately, the CMSSM suffers from the so-called $\mu$
problem~\cite{Kim:1983dt}. This problem is elegantly solved in the
framework of the next-to-minimal supersymmetric standard model
(NMSSM)~\cite{Nilles:1982dy}. In addition to the MSSM Yukawa
couplings for quarks and leptons, the NMSSM superpotential contains
two additional terms involving the Higgs doublet superfields, $H_1$
and $H_2$, and the new superfield $S$, a singlet under the SM gauge
group $SU(3)_c \times SU(2)_L \times U(1)_Y$~\cite{Cerdeno:2004xw}
\begin{equation}\label{2:Wnmssm}
W= \epsilon_{ij} \left( Y_u \, H_2^j\, Q^i \, u + Y_d \, H_1^i\, Q^j
\, d + Y_e \, H_1^i\, L^j \, e \right) - \epsilon_{ij} \lambda \,S
\,H_1^i H_2^j +\frac{1}{3} \kappa S^3\,
\end{equation}
where we take $H_1^T=(H_1^0, H_1^-)$, $H_2^T=(H_2^+, H_2^0)$, $i,j$ are
$SU(2)$ indices, and $\epsilon_{12}=1$. In this model, the usual MSSM bilinear
$\mu$ term is absent from the superpotential, and only dimensionless trilinear
couplings are present in $W$. However, when the scalar component of $S$ acquires
a VEV, an effective interaction $\mu H_1 H_2$ is generated, with $\mu \equiv
\lambda \langle S \rangle$.

Finally, the soft SUSY
breaking terms are given by~\cite{Cerdeno:2004xw}
\begin{align}\label{2:Vsoft}
-\mathcal{L}_{\text{soft}}=&\,
 {m^2_{\tilde{Q}}} \, \tilde{Q}^* \, \tilde{Q}
+{m^2_{\tilde{U}}} \, \tilde{u}^* \, \tilde{u}
+{m^2_{\tilde{D}}} \, \tilde{d}^* \, \tilde{d}
+{m^2_{\tilde{L}}} \, \tilde{L}^* \, \tilde{L}
+{m^2_{\tilde{E}}} \, \tilde{e}^* \, \tilde{e}
 \nonumber \\
&
+m_{H_1}^2 \,H_1^*\,H_1 + m_{H_2}^2 \,H_2^* H_2 +
m_{S}^2 \,S^* S \nonumber \\
&
+\epsilon_{ij}\, \left(
A_u \, Y_u \, H_2^j \, \tilde{Q}^i \, \tilde{u} +
A_d \, Y_d \, H_1^i \, \tilde{Q}^j \, \tilde{d} +
A_e \, Y_e \, H_1^i \, \tilde{L}^j \, \tilde{e} + \text{H.c.}
\right) \nonumber \\
&
+ \left( -\epsilon_{ij} \lambda\, A_\lambda S H_1^i H_2^j +
\frac{1}{3} \kappa \,A_\kappa\,S^3 + \text{H.c.} \right)\nonumber \\
& - \frac{1}{2}\, \left(M_3\, \lambda_3\, \lambda_3+M_2\,
\lambda_2\, \lambda_2 +M_1\, \lambda_1 \, \lambda_1 + \text{H.c.}
\right) \,
\end{align}

Clearly, the NMSSM is very similar to the MSSM. Despite the similarities
between the two particle physics models, the Higgs sector as well as the neutralino
mass matrix and mass eigenstates in the NMSSM are more complicated and richer
compared to the corresponding ones in the MSSM.

In particular, in the Higgs sector we have now two CP-odd neutral,
and three CP-even neutral Higgses. We make the assumption that there
is no CP-violation in the Higgs sector at tree level, and neglecting 
loop level effects, the CP-even
and CP-odd states do not mix. We are not interested in the CP-odd
states, while the CP-even Higgs interaction and physical eigenstates
are related by the transformation
\begin{equation}\label{2:Smatrix}
h_a^0 = S_{ab} H^0_b\,
\end{equation}
where $S$ is the unitary matrix that diagonalises the CP-even
symmetric mass matrix, $a,b = 1,2,3$, and the physical eigenstates
are ordered as $m_{h_1^0} \lesssim
m_{h_2^0} \lesssim m_{h_3^0}$.

In the neutralino sector the situation is again more involved, since
the fermionic component of $S$ mixes with the neutral Higgsinos,
giving rise to a fifth neutralino state. In the weak interaction
basis defined by ${\Psi^0}^T \equiv \left(\tilde B^0=-i
\lambda^\prime, \tilde W_3^0=-i \lambda_3, \tilde H_1^0, \tilde
H_2^0, \tilde S \right)\,$, the neutralino mass terms in the
Lagrangian are~\cite{Cerdeno:2004xw}
\begin{equation}
\mathcal{L}_{\mathrm{mass}}^{\tilde \chi^0} =
-\frac{1}{2} (\Psi^0)^T \mathcal{M}_{\tilde \chi^0} \Psi^0 + \mathrm{H.c.}\,,
\end{equation}
with $\mathcal{M}_{\tilde \chi^0}$ a $5 \times 5$ matrix,
{\footnotesize \begin{equation}
  \mathcal{M}_{\tilde \chi^0} = \left(
    \begin{array}{ccccc}
      M_1 & 0 & -M_Z \sin \theta_W \cos \beta &
      M_Z \sin \theta_W \sin \beta & 0 \\
      0 & M_2 & M_Z \cos \theta_W \cos \beta &
      -M_Z \cos \theta_W \sin \beta & 0 \\
      -M_Z \sin \theta_W \cos \beta &
      M_Z \cos \theta_W \cos \beta &
      0 & -\lambda s & -\lambda v_2 \\
      M_Z \sin \theta_W \sin \beta &
      -M_Z \cos \theta_W \sin \beta &
      -\lambda s &0 & -\lambda v_1 \\
      0 & 0 & -\lambda v_2 & -\lambda v_1 & 2 \kappa s
    \end{array} \right)
  \label{neumatrix}
\end{equation}}
The above matrix can be diagonalised by means of a unitary matrix
$N$
\begin{equation}
N^* \mathcal{M}_{\tilde \chi^0} N^{-1} = \operatorname{diag}
(m_{\tilde \chi^0_1}, m_{\tilde \chi^0_2}, m_{\tilde \chi^0_3},
m_{\tilde \chi^0_4}, m_{\tilde \chi^0_5})\,
\end{equation}
where $m_{\tilde \chi^0_1}$ is the lightest
neutralino mass. Under the above assumptions, the lightest neutralino can be
expressed as the combination
\begin{equation} \label{composition}
\tilde \chi^0_1 = N_{11} \tilde B^0 + N_{12} \tilde W_3^0 + N_{13}
\tilde H_1^0 + N_{14} \tilde H_2^0 + N_{15} \tilde S\,
\end{equation}
In the following, neutralinos with $N^2_{11}>0.9$, or
$N^2_{15}>0.9$, will be referred to as bino- or singlino-like,
respectively.

Similarly to the CMSSM, in the constrained next-to-minimal supersymmetric standard model
the universality of $m_0, A_0, m_{1/2}$ at the GUT scale is again assumed,
with the only parameters now being~\cite{Hugonie:2007vd} \\
\centerline{$tan \beta, m_0, A_0, m_{1/2}, \lambda, A_k$} and the
sign of the $\mu$ parameter can be chosen at will.

We end the discussion on the particle physics model here, by making
a final remark regarding the differences between the CMSSM and the
cNMSSM. In the CMSSM the lightest neutralino is mainly a bino in
most of the parameter space, and low values of $m_0$ are disfavored
because they lead to charged sleptons that are lighter than the
neutralino $\chi_1^0$, while in the cNMSSM the lightest neutralino
is mainly a singlino in large regions of the parameter space, thanks
to which the charged LSP problem can be
avoided~\cite{Hugonie:2007vd}. Furthermore, in the cNMSSM there are
more mechanisms that contribute to the neutralino relic
density~\cite{Hugonie:2007vd}.

\subsection{Gravitino production}

In the usual case (not in the degenerate scenario) where the neutralino decays into a gravitino 
and standard model particles with a lifetime typically in the range $(10^4-10^8)\: sec$,
for the gravitino abundance we take the relevant production mechanisms into account
and impose the cold
dark matter constraint~\cite{Komatsu:2008hk}
\begin{equation}
0.1097 < \Omega_{cdm} h^2=\Omega_{3/2} h^2 < 0.1165
\end{equation}
At this point it is convenient to define the gravitino yield, $Y_{3/2} \equiv n_{3/2}/s$,
where $n_{3/2}$ is
the gravitino number density, $s=h_{eff}(T) \frac{2 \pi^2}{45} T^3$ is the entropy density
for a relativistic
thermal bath, and $h_{eff}$ counts the relativistic degrees of freedom. The gravitino
abundance $\Omega_{3/2}$
in terms of the gravitino yield is given by
\begin{equation}
\Omega_{3/2} h^2=\frac{\mgravitino s(T_0) Y_{3/2} h^2}{\rho_{cr}}=2.75 \times 10^{8} \left ( \frac{\mgravitino}{\textrm{GeV}}
\right ) Y_{3/2}(T_0)
\end{equation}
where we have used the values
\begin{eqnarray}
T_0 & = & 2.73 K=2.35 \times 10^{-13}~\textrm{GeV} \\
h_{eff}(T_0)& = & 3.91 \\
\rho_{cr}/h^2& = & 8.1 \times 10^{-47}~\textrm{GeV}^4
\end{eqnarray}
The total gravitino yield has two contributions, namely one from the thermal bath, and one
from the out-of-equillibrium NLSP decay.
\begin{equation}
Y_{3/2}=Y_{3/2}^{TP}+Y_{3/2}^{NLSP}
\end{equation}
The contribution from the thermal production has been computed in
~\cite{Bolz:2000fu,Pradler:2006qh,Rychkov:2007uq}. In~\cite{Bolz:2000fu} the gravitino
production was
computed in leading order in the gauge coupling $g_3$,
in~\cite{Pradler:2006qh} the thermal
rate was
computed in leading order in all Standard Model gauge couplings $g_Y, g_2, g_3$, and
in~\cite{Rychkov:2007uq}
new effects were taken into account, namely: a) gravitino production via gluon $\rightarrow$
gluino
$+$ gravitino and other decays, apart from the previously considered $2 \rightarrow 2$
gauge scatterings,
b) the effect of the top Yukawa coupling, and c) a proper treatment of the reheating
process.
Here we
shall use the result of~\cite{Bolz:2000fu} since the corrections
of~\cite{Pradler:2006qh,Rychkov:2007uq}
do not alter our conclusions. Therefore the thermal gravitino production is given by
\begin{equation}
Y_{3/2}^{TP}=0.29 \times 10^{-12} \: \left (
\frac{T_R}{10^{10}~\textrm{GeV}}\right ) \: \left (1+\frac{1}{3}
\frac{m_{\tilde{g}}^2}{\mgravitino^2}\right )
\end{equation}
or, approximately for a light gravitino, $\mgravitino \ll m_{\tilde{g}}$
\begin{equation}
Y_{3/2}^{TP} \simeq 10^{-13} \: \left ( \frac{T_R}{10^{10}~\textrm{GeV}}
\right ) \: \left ( \frac{m_{\tilde{g}}}{\mgravitino} \right )^2
\end{equation}
with $\mgravitino$ the gravitino mass and $m_{\tilde{g}}$ the gluino mass.

The second contribution to the gravitino abundance comes from the decay of the NLSP
\begin{equation}
\Omega_{3/2}^{NLSP} h^2 = \frac{\mgravitino}{m_{NLSP}} \: \Omega_{NLSP} h^2
\end{equation}
with $m_{NLSP}$ the mass of the NLSP, and $\Omega_{NLSP} h^2$ the
abundance the NLSP would have, had it not decayed into the gravitino. In the
limit where $m_{NLSP} \rightarrow \mgravitino$ and $\tau_{NLSP} \gg 10^{17}~sec$
the scenario looks as if one would have a two-component dark matter with the NLSP
contribution $\Omega_{NLSP}h^2$, and a gravitino contribution from thermal production
only, $Y_{3/2}^{TP}$. Therefore, in the degenerate scenario with 
$m_{NLSP} \simeq \mgravitino$ the WMAP bound becomes
\begin{equation}
0.1097 < \Omega_{cdm} h^2=\Omega_{NLSP} h^2+\Omega_{3/2}^{TP}h^2 < 0.1165
\end{equation}
where from now on the NLSP is the lightest neutralino, $\chi=NLSP$.

\section{Constraints and results}

- Spectrum and collider constraints: We have used
NMSSMTools~\cite{Ellwanger:2004xm}, a computer software that
computes the masses of the Higgses and the superpartners, the
couplings, and the relic density of the neutralino, for a given set
of the free parameters. We have performed a random scan in the whole
parameter space (with fixed $\mu > 0$ motivated by the muon
anomalous magnetic moment), and we have selected only those points
that satisfy i) theoretical requirements, such as neutralino LSP,
correct electroweak symmetry breaking, absence of tachyonic masses
etc., and ii) LEP bounds on the Higgs mass, collider bounds on SUSY
particle masses, and experimental data from
B-physics~\cite{precision, Yao:2006px}. For all these good points
the lightest neutralino is either a bino or a singlino, and contrary
to the case where neutralino is the dark matter particle, here we do
not require that the neutralino relic density falls within
the allowed WMAP range.

- As we have already mentioned, the total dark matter abundance, and
not the neutralino one, should satisfy the cold dark matter
constraint~\cite{Komatsu:2008hk}
\begin{equation}
0.1097 < \Omega_{cdm} h^2=\Omega_{\chi} h^2+\Omega_{3/2}^{TP} h^2 < 0.1165
\end{equation}
that relates the reheating temperature after inflation to the
gravitino mass as follows
\begin{equation}
0.11 = A(\mgravitino, m_{\tilde{g}}) T_R+\Omega_{\chi} h^2
\end{equation}
For a given point in the cNMSSM parameter space, the complete
spectrum and couplings have been computed, and we are left with two
more free parameters, namely the gravitino mass and the reheating
temperature after inflation. The gravitino mass is equal essentially to the
neutralino mass, and the precise value can be determined if we specify the neutralino
lifetime. In the discussion to follow we have used a neutralino lifetime 
$\tau=10^{26} \: sec$, although the results
are not sensitive to it, and the figures we have produced for different values of the 
lifetime cannot be distinguished. Finally, the reheating temperature after inflation is 
obtained from the cold dark matter constraint. The thermal production contribution cannot 
be larger than the total dark matter abundance, and for this we can
already obtain an upper bound on the reheating temperature
\begin{equation}
T_R \leq 4.1 \times 10^9 \left ( \frac{\mgravitino}{100~\textrm{GeV}} \right ) \: \left
( \frac{\textrm{TeV}}{m_{\tilde{g}}} \right )^2 \: \textrm{GeV}
\end{equation}
Assuming a gluino mass $m_{\tilde{g}} \sim 1$~TeV, we can see that for a
heavy gravitino, $\mgravitino \sim 100$~GeV, it is possible to obtain a reheating
temperature large enough for thermal leptogenesis.

- For neutralino NLSP in the degenerate scenario, the only decay mode is 
$\chi \to \gamma \gravitino$,
for which the decay width can be computed once the supergravity Largrangian is 
known~\cite{Cremmer:1982en}, and it is given by~\cite{Feng:2004mt, Ellis:2003dn}
\begin{equation}
\Gamma(\chi \to \gamma \gravitino)
=
\frac{| N_{11} \cos \theta_W + N_{12} \sin \theta_W |^2}
{48\pi M_*^2} \
\frac{m_{\chi}^5}{m_{\gravitino}^2}
\left[1 - \frac{m_{\gravitino}^2}{m_{\chi}^2} \right]^3
\left[1 + 3 \frac{m_{\gravitino}^2}{m_{\chi}^2} \right]
\label{neutralinogamma}
\end{equation}
where $M_*$ is the Planck mass, $m_{\chi}$ is the neutralino mass,
and $\theta_W$ is the weak angle. In the limit where the mass difference
$\Delta m \equiv m_\chi-\mgravitino$ is much lower than the masses themselves,
$\Delta m \ll m_\chi, \mgravitino$, the neutralino lifetime becomes
\begin{equation}
\tau=\frac{1.78 \times 10^{13}~sec}{|N_{11} \cos \theta_W + N_{12} \sin \theta_W|^2} \: 
\left( \frac{GeV}{\Delta m} \right )^3
\end{equation}
From this formula one can see that for a mostly bino-neutralino a mass difference of 1 MeV 
is already enough to give a neutralino lifetime larger than the age of the universe.

- Neutralino-Nucleon spin-independent cross-section: LHC is now running and collecting data.
Although LHC is a powerful machine to look for physics beyond the standard model, it is 
known that other facilities are also needed to offer
complementary information towards the direction of searching for supersymmetry and 
identifying dark matter. The gravitino interactions are suppressed by the Planck mass,
and therefore direct production of gravitinos at colliders and/or direct detection prospects
seem to be hopeless. On the other hand, for a weakly interacting neutralino there are existing 
as well as future experiments that put experimental limits on the nucleon-neutralino 
cross-section. The spin-independent cross-section is given by
\begin{equation}
\sigma_{\chi-N}=\frac{4 m^2_r}{\pi} \, f_N^2\,
\end{equation}
where $m_r$ is the Nucleon-neutralino reduced mass,
$m_r=m_N m_{\chi}/(m_N +m_{\chi})$,
and
\begin{equation}
\frac{f_N}{m_N}\,=\,
\operatornamewithlimits{\sum}_{q=u,d,s} f_{Tq}^{(N)}
\frac{\alpha_{q}}{m_q} + \frac{2}{27}  f_{TG}^{(N)}
\operatornamewithlimits{\sum}_{q=c,b,t}
\frac{\alpha_{3q}}{m_q} \,
\end{equation}
In the above, $f_{TG}^{(N)}= 1- \operatornamewithlimits{\sum}_{q=u,d,s}
f_{Tq}^{(N)}$, we have taken the following
values for the hadronic matrix elements~\cite{Ellis:2000ds}:
\begin{align}\label{3:hadronvalues}
f_{Tu}^{(p)}= 0.020 \pm 0.004\,, \ \ \ &
f_{Td}^{(p)}= 0.026 \pm 0.005\,, &
f_{Ts}^{(p)}= 0.118 \pm 0.062\,,
\nonumber \\
f_{Tu}^{(n)}= 0.014 \pm 0.003\,, \ \ \ &
f_{Td}^{(n)}= 0.036 \pm 0.008\,, &
f_{Ts}^{(n)}= 0.118 \pm 0.062\,.
\end{align}
and $\alpha_q$ is the coupling in the effective Lagrangian
\begin{equation}
\mathcal{L}_{\mathrm{eff}} =\alpha_{i} \,\bar{\chi} \,\chi \, \bar q_i\, q_i
\end{equation}
where $i=1,2$ denotes up- and down-type quarks, and the Lagrangian is
summed over the three quark generations. The coupling $\alpha_q$ can be decomposed
into two parts, $\alpha_q=\alpha_q^h+\alpha_q^{\tilde{q}}$, where the first term
is the t-channel exchange of a neutral Higgs (Fig.~1), while the second term is the s-channel
exchange of a squark (Fig.~2). The expressions for $\alpha_q$ is terms of the masses and 
couplings of the model can be found in~\cite{Cerdeno:2004xw}.

Our main results are summarized in the figures below. In Fig.~3 and Fig.~4 we show the 
Nucleon-neutralino spin-independent cross section (in $\textrm{cm}^2$) versus neutralino mass and 
lightest Higgs boson (in GeV) respectively. The blue region corresponds to a bino neutralino, 
while the green
region corresponds to a singlino neutralino, and the curves are the current experimental limits
from CDMS~\cite{Ahmed:2009zw}. According to our results the bino scenario is already ruled 
out, while in the singlino case the 
upper region can be probed by future experiments. In Fig.~5 we show the reheating temperature after
inflation as a function of the neutralino/gravitino mass. The blue region corresponds to a bino,
the blue points correspond to singlino, and finally the red points correspond to singlino with 
relatively high values of the cross-section, namely $\sigma_{\chi-N} > 10^{-47}~cm^2$. The largest
values of $T_R$ correspond to a bino, which is ruled out, and for the singlino with relatively
high values of cross-section we obtain a reheating temperature $T_R \simeq 5 \times 10^9 \: GeV$
for a neutralino/gravitino mass $m_{\chi} \simeq \mgravitino \simeq 200 \: GeV$. In the last
figure we show the ($m_0$-$m_{1/2}$) plane ($m_0$ and $m_{1/2}$ in GeV) for singlino points 
with a cross-section larger than $10^{-47}~cm^2$, or lower than $10^{-47}~cm^2$. We see that
$m_0$ is not larger than 600 GeV, and therefore future direct detection experiments cannot
probe a region of the parameter space which can neither be probed by LHC.

Before ending the discussion, let us briefly comment on a possible collider signature of our model. 
In the singlino-like neutralino case, where the coupling
$\lambda$ is small, $\lambda \leq 0.01$, the lightest Higgs can be very light, $m_h < 60$~GeV, 
which has a significant singlet composition,
thus escaping detection and being in agreement with accelerator data. In this
case the next-to-lightest Higgs is the SM-like Higgs, with a mass $m_H \simeq (116-118)$~GeV, 
and the decay channel $H \rightarrow hh$ is kinematically allowed. Since the lightest Higgs
is expected to exit the detector without been seen, the decay channel $H \rightarrow hh$ 
is an invisible one. This is to be contrasted
with the cases of SM and of the MSSM, where the Higgs (in the SM) and the 
SM-like Higgs (in the MSSM) with a mass in the above range decays (almost entirely) into 
$b \bar{b}$ and $\tau^+ \tau^-$, with the sum of the two branching ratios being practically 
one~\cite{Djouadi:2005gi}.
In Table~1 we show the range of the parameters of the model where we obtain a very light Higgs
and the decay channel $H \rightarrow hh$ is kinematically allowed. For most of the points
the branching ratio is negligible, even as low as $\sim 10^{-9}$, but points exist for 
which the branching ratio becomes sizable, $BR \sim 0.1$,
with the maximum value obtained being $BR(H \rightarrow hh) = 0.13$. In Table~2 we show the Higgs
boson masses $M_H, M_h$ and the branching ratio for four points.

\begin{table}
\begin{center}
\begin{tabular}{|c|c|c|c|c|c|}
\hline
$ \lambda $ & $ tan \beta $ & $A_0$~(GeV) & $A_k$~(GeV) & $m_0$~(GeV) & $ M_{1/2}(GeV) $  \\
\hline
0.0102 & 34.89 & -127.24 & -106.41 & 124.28 & 606.5   \\
\hline
0.000127 & 25.49 & -225.09 & -170.8 & 10.6 & 401.43  \\                       
\hline
\end{tabular}
\caption{Range of the parameters of the model where the decay channel 
$H \rightarrow hh$ is kinematically allowed. The last row shows the minimum value, while the 
row in the middle shows the maximum value of the parameters.}
\end{center}
\end{table}

\begin{table}
\hspace{-2.5cm}
\begin{tabular}{|c|c|c|c|c|c|c|c|c|c|}
\hline
$\lambda$ & $tan \beta$ & $A_0$~(GeV) & $A_k$~(GeV) & $m_0$~(GeV) & $M_{1/2}$~(GeV) & $M_H$~(GeV) & $M_h$~(GeV) & $BR(H \rightarrow hh)$ \\
\hline
 0.000107 & 34.69 & -202.66 & -115.23 & 22.65 & 581.2 & 116.97  & 54.81 & $3.4 \times 10^{-9}$ \\
\hline
 0.00216  & 30.25 & -158.28 & -135.98 & 49.33 & 467.87 & 115.8  & 46.8 & $2.82 \times 10^{-4}$ \\                       
\hline
 0.009154 & 32.3  & -181.5  & -131.01 & 16.2  & 540.57 & 118.33 & 49.44 &  0.13  \\
\hline
 0.005064 & 30.94 & -185.46 & -169.65 & 35.86 & 552.87 & 117.29 & 57.17 &   0.01  \\
\hline
\end{tabular}
\caption{Higgs boson masses, the branching ratio and the values of the parameters for four points.}
\end{table}

\section{Conclusions}

In the framework of NMSSM, which solves the $\mu$ problem, we have assumed that the
gravitino LSP and the lightest neutralino NLSP are degenerate in mass.
Under this assumption the neutralino becomes extremely long-lived 
avoiding the BBN bounds. In this scenario we have a two component dark matter made
out of the absolutely stable gravitino and the quasi-stable neutralino. We have performed
a random scan over the whole parameter space keeping the points that satisfy the available
collider constraints plus the WMAP bound for dark matter. These points
correspond to either a bino or a singlino neutralino. We have computed the 
neutralino-nucleon spin-independent cross section as a function of the neutralino mass
and the lightest Higgs mass, and we find that the bino case is ruled out (see Fig.~3 and
Fig.~4). Then we explored the ($m_0-m_{1/2}$) parameter space, and the reheating temperature 
dependence of the neutralino/gravitino mass for the singlino points that correspond to 
cross section values to be probed by future experiments. Finally, we have briefly discussed
an interesting possibility for collider signatures, namely the possibility of having an
invisible decay channel $H \rightarrow hh$, where $H$ is the SM-like Higgs and $h$ is the
lightest Higgs that escapes detection, with a sizable branching ratio and maximum allowed
value $BR(H \rightarrow hh) \simeq 0.13$.

\section*{Acknowledgments}

We are greatful to L.~Boubekeur for reading the manuscript, and for useful comments 
and discussions. The authors acknowledge financial
support from spanish MEC and FEDER (EC) under grant FPA2008-02878,
and Generalitat Valenciana under the grant PROMETEO/2008/004.

\newpage


\begin{figure}
\centerline{\epsfig{figure=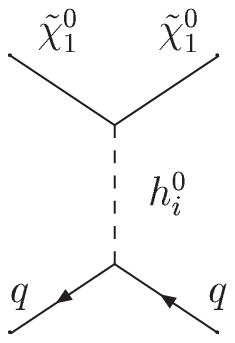,height=5cm,angle=0}}
\caption{Neutralino scattering off a nucleon by a neutral Higgs boson exchange.}
\end{figure}

\begin{figure}
\centerline{\epsfig{figure=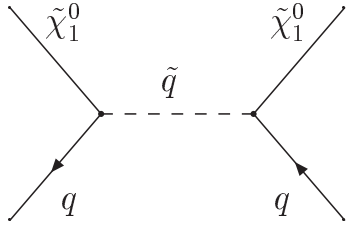,height=3cm,angle=0}}
\caption{Neutralino scattering off a nucleon by a squark exchange.}
\end{figure}

\newpage

\begin{figure}
\centerline{\epsfig{figure=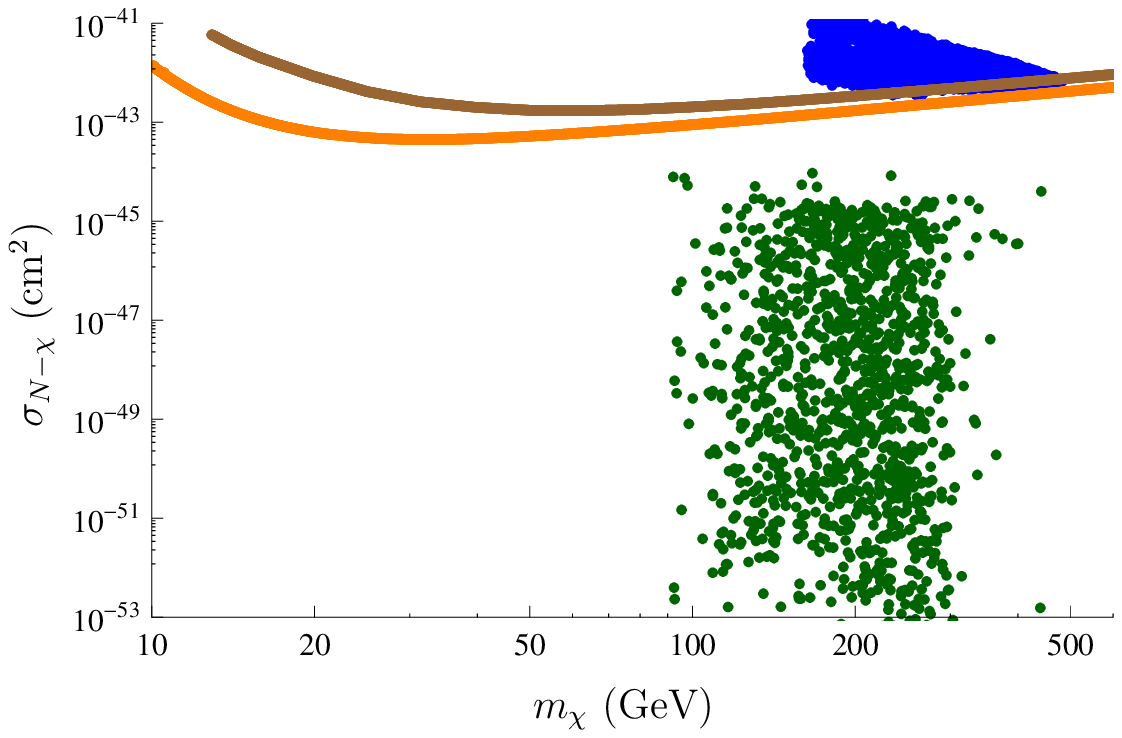,height=8cm,angle=0}}
\caption{Spin-independent neutralino-nucleon (proton) cross-section versus 
neutralino mass. Shown are the available experimental bounds from CDMS, and the 
predictions of the theoretical model. The blue region corresponds to a bino-like
neutralino, while the green points correspond to a singlino-like neutralino.}
\end{figure}

\begin{figure}
\centerline{\epsfig{figure=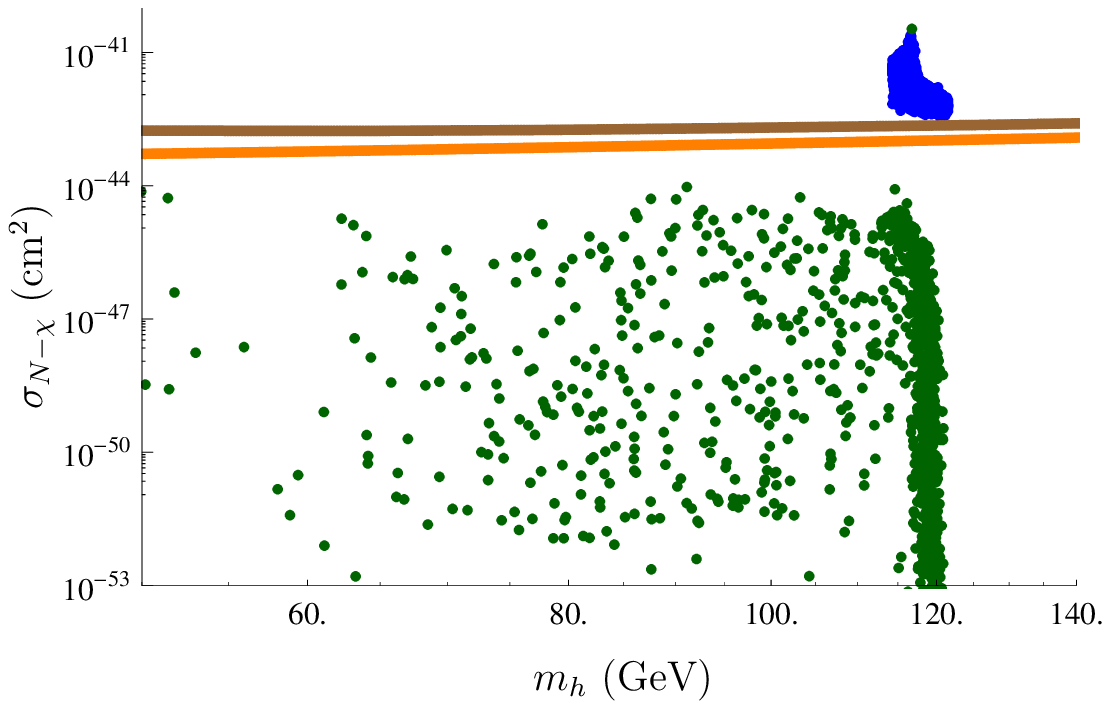,height=8cm,angle=0}}
\caption{Spin-independent neutralino-nucleon (proton) cross-section versus 
the lightest Higgs mass. Shown are the available experimental bounds CDMS, and 
the predictions of the theoretical model. The blue region corresponds to a bino-like
neutralino, while the green points correspond to a singlino-like neutralino.}
\end{figure}

\begin{figure}
\centerline{\epsfig{figure=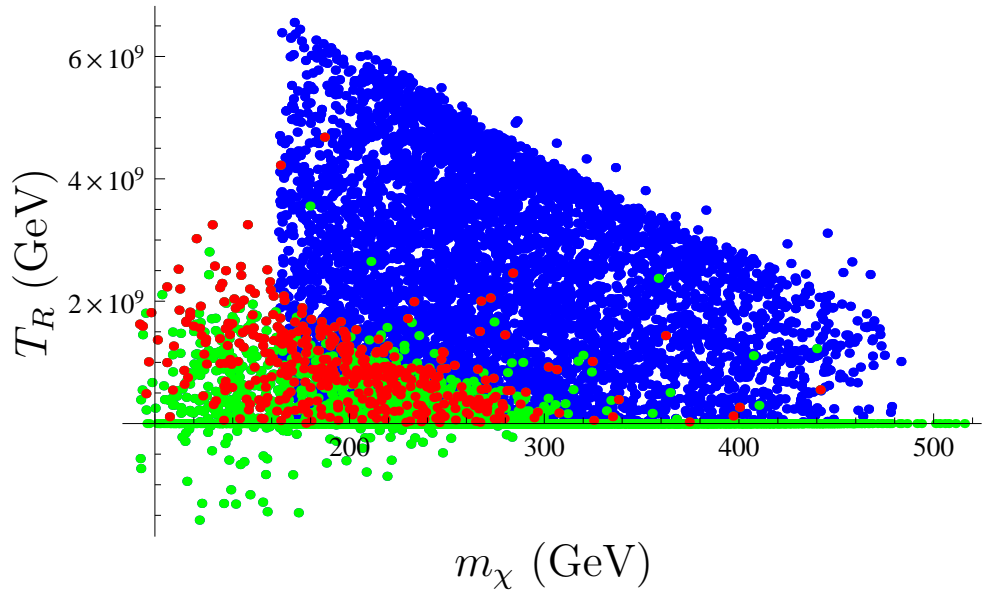,height=8cm,angle=0}}
\caption{Reheating temperature versus neutralino (or gravitino mass). Blue points
correspond to bino, green points correspond to singlino, and red points correspond
to singlino with a cross-section larger than $10^{-47}~cm^2$.}
\end{figure}

\begin{figure}
\centerline{\epsfig{figure=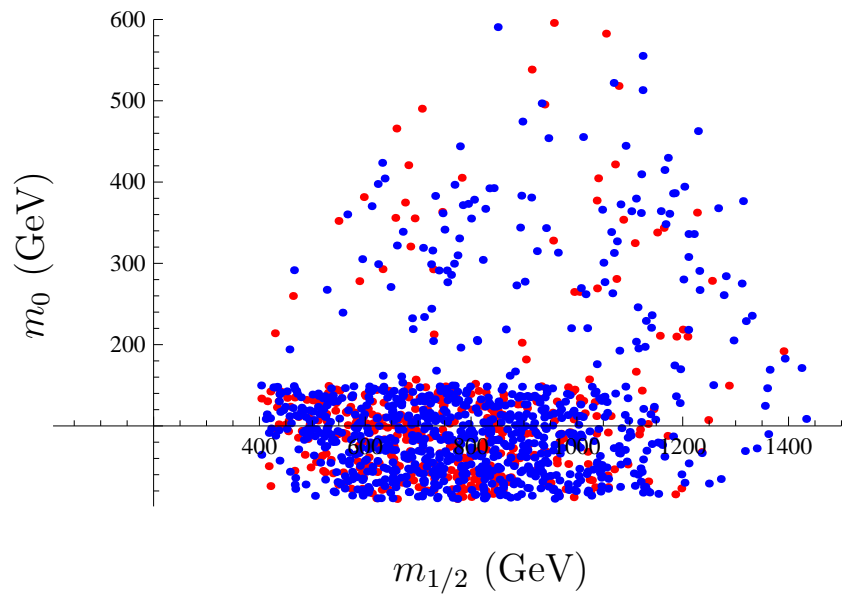,height=8cm,angle=0}}
\caption{The ($m_0$-$m_{1/2}$) plane for the singlino points. One color corresponds to
a cross-section larger than $10^{-47}~cm^2$, and the other color corresponds to
a cross-section lower than $10^{-47}~cm^2$.}
\end{figure}

\end{document}